\documentclass[pra,twocolumn,showpacs,preprintnumbers,superscriptaddress]
{revtex4}

\usepackage{times}
\usepackage{bm}
\usepackage{float}
\usepackage{graphicx}
\usepackage{amsbsy}
\usepackage{amsmath}
\usepackage{amsfonts}
\usepackage{amsthm}
\begin{document}
	
\theoremstyle{plain}
\newtheorem{theorem}{Theorem}
\newtheorem{lemma}[theorem]{Lemma}
\newtheorem{corollary}[theorem]{Corollary}
\newtheorem{proposition}[theorem]{Proposition}
\newtheorem{conjecture}[theorem]{Conjecture}

\theoremstyle{definition}
\newtheorem{definition}[theorem]{Definition}

\theoremstyle{remark}
	\newtheorem*{remark}{Remark}
	\newtheorem{example}{Example}
	\title{Detection and Classification of Three-qubit States Using $l_{1}$ Norm of Coherence}
	\author{Anu Kumari, Satyabrata Adhikari}
	\email{mkumari_phd2k18@dtu.ac.in, satyabrata@dtu.ac.in} \affiliation{Delhi Technological
		University, Delhi-110042, Delhi, India}
	
\begin{abstract}
	\begin{center}
		\textbf{Abstract}
	\end{center}
Entanglement is a purely quantum mechanical phenomenon and thus it has no classical analog. On the other hand, coherence is a well-known phenomenon in classical optics and in quantum mechanics. Recent research shows that quantum coherence may act as a useful resource in quantum information theory. We will employ here quantum coherence to detect and classify the entanglement property of three-qubit states. Moreover, we have shown that if any three-qubit state violates another necessary condition for the detection of a general biseparable state then the given three-qubit state cannot be a biseparable state. Since there are only three categories of states for the three-qubit system so if we detect that the state under probe is neither a separable nor a biseparable state then we can definitely conclude that the given three-qubit state is a genuine entangled state. We have illustrated our results with a few examples.
\end{abstract}
\pacs{03.67.Hk, 03.67.-a} \maketitle
	
\section{Introduction}
Quantum entanglement and coherence are two fundamental features that arise from the superposition principle of quantum mechanics. One notable difference between these two features is that coherence may exist in a single system while more than one system or more than one degree of freedom is required for entanglement. Secondly, the non-zero off-diagonal elements in the density matrix signify the presence of coherence in the quantum system while it may not ensure the existence of entanglement in the given composite quantum system. Both quantum entanglement and quantum coherence can be used as a resource \cite{horodecki4,streltsov}. M. Hillery has shown that coherence may act as a useful resource in the Deutsch-Jozsa algorithm \cite{hillery}. On the other hand, quantum entanglement has many applications in quantum information processing tasks, notably, quantum teleportation \cite{bennett1}, quantum superdense coding \cite{bennett2}, quantum remote state preparation \cite{pati}, quantum cryptography \cite{gisin} etc. Extensive research has already been carried out to understand the non-classical feature of bipartite system \cite{peres,horodecki3,wootters,vidal,plenio,horodecki4,zhang2}. Eventually, when we increase the number of parties in the system, the complications in the shared system between the parties also increase. Therefore, it is indispensable to understand the entanglement properties of the shared multipartite system.\\
In this work, we will consider one of the most important problems in quantum information theory i.e. detection and classification of multipartite entanglement. We have studied this problem specifically in the case of a tripartite system and mainly discuss the detection and classification of a three-qubit biseparable state. We will derive coherence-based inequalities for the detection of three-qubit biseparable systems.\\
Let us start our discussion on a tripartite system by assuming that the Hilbert spaces $H_{A}$, $H_{B}$ and $H_{C}$, which describe each subsystem $A$, $B$, $C$ of a tripartite system, is spanned by the computational basis states $|0\rangle$ and $|1\rangle$ respectively.\\
Any three-qubit state may be classified as fully separable states, biseparable states, or genuinely entangled states \cite{dur4}. If the three-qubit state is shared by three distant parties Alice, Bob, and Charlie then the shared state is either a fully separable or biseparable state or a genuine entangled state. The fully separable and biseparable state may be expressed in the form as \cite{dur3,toth2}
\begin{eqnarray}
\rho_{sep}^{ABC}=\sum_{i}p_{i}\rho_{i}^{A}\otimes\rho_{i}^{B}\otimes\rho_{i}^{C},\sum_{i}p_{i}=1
\label{generalseparable}
\end{eqnarray}
\begin{eqnarray}
\rho_{bisep} = p_1 \rho^{A-BC}_{bisep} + p_2 \rho^{B-AC}_{bisep} + p_3 \rho^{C-AB}_{bisep},\sum_{i}p_{i}=1
\label{generalbiseparable}
\end{eqnarray}
where,
\begin{eqnarray}
\rho_{bisep}^{A-BC}=\sum_{i}|a_i\rangle_A \langle a_{i}|\otimes |\phi_{i}\rangle_{BC}\langle \phi_{i}|\nonumber\\
\rho_{bisep}^{B-AC}=\sum_{i}|b_i\rangle_B \langle b_{i}|\otimes |\phi_{i}\rangle_{AC}\langle \phi_{i}|\nonumber\\
\rho_{bisep}^{C-AB}=\sum_{i}|c_i\rangle_C \langle c_{i}|\otimes |\phi_{i}\rangle_{AB}\langle \phi_{i}|\nonumber
\end{eqnarray}
Here, $|a_{i}\rangle $, $|b_{i}\rangle $ and $|c_{i}\rangle $  are (unnormalized) states of systems
A, B and C, respectively and $|\phi_{i}\rangle $ are states of two systems and $0\leq p_i\leq 1$ in equations (\ref{generalseparable}) and (\ref{generalbiseparable}).\\
The density operators $\rho_{i}^{x},~x=A, B, C$ lying on the Hilbert spaces of dimension 2 whereas the density operators $\rho_{i}^{yz},~y,z=A, B, C$ lying on the composite Hilbert space of two qubits. If any three-qubit state does not fall under the above forms given by (\ref{generalseparable}) and (\ref{generalbiseparable}) then the state is a genuine three-qubit entangled state.\\
It has been observed that the three-qubit biseparable states are important and useful in many contexts. The importance of biseparable states comes from the fact that they may be used as unextendible biseparable bases (UBB) which are proved to be useful to construct genuinely entangled subspaces \cite{agrawal}. Instead of a genuine tripartite entangled state, it has been shown that a biseparable state is enough to use in the controlled quantum teleportation protocol as a resource state \cite{barasinski}. Barasinski et.al. \cite{artur2}  have analyzed the fidelity of the controlled quantum teleportation via mixed biseparable state and have concluded that a statistical mixture of biseparable states can be suitable for the perfect controlled quantum teleportation. Further, it has been shown that there exists a special class of biseparable state i.e. a non-maximally entangled mixed biseparable X state, which can be useful as a resource state for the attainment of high fidelity in controlled quantum teleportation \cite{paulson}. Recently, it is shown that the biseparable states can also be used in obtaining the non-zero conference key \cite{carrara}.\\
Since the three-qubit biseparable state has potential applications in quantum information theory, it is crucial to detect the three-qubit biseparable state. There are a lot of earlier works in this line of research but we mention here a few of them \cite{eggeling,zhaofei,bancal,zhao12,lohmayer,salay,novo,shen,das,han,datta1,kairon,satyabrata}. Eggeling and Werner \cite{eggeling} provided the necessary and sufficient criteria in terms of the projection parameters to detect the biseparable state. But this result is true for only one bipartition cut $A-BC$. In \cite{zhaofei}, it has been shown that the witness operator may be constructed to distinguish fully separable state, biseparable state, and genuine entangled states for a multipartite system. An n-partite inequality is presented in \cite{bancal}, whose violation by a state implies that the state under investigation is not a biseparable state. A non-linear entanglement witness operator has been constructed to identify all three types of three-qubit states \cite{zhao12}. Using entanglement measures biseparability in mixed three-qubit systems has also been analyzed in detail in  \cite{lohmayer}-\cite{salay}.
The necessary and sufficient condition for the detection of permutationally invariant three-qubit biseparable has been studied in \cite{novo}. The multipartite biseparable entangled states under any bipartite partitions can be detected using linear contraction methods \cite{shen}. A set of Bell inequalities was introduced in \cite{das}, which can distinguish separable, biseparable, and genuinely entangled pure three-qubit states. The method of construction of biseparable state is given in \cite{han} and the classification of three-qubit pure states has been studied in \cite{datta1}. Recently, we have discriminated a particular class of three-qubit GHZ and W class of states using coherence-based inequality \cite{kairon}.\\
The whole work is organized in the following way: In Sec. II, we obtain the expression of $l_1$ norm of coherence of the tensor product of two quantum states which forms two subsystems of an n-partite system. 
Furthermore, we derive coherence-based inequality for the detection of a particular form of a three-qubit biseparable state. At the end of the section, we support our detection criterion with a few examples. In Sec. III, we have derived coherence-based inequalities for the detection of general three-qubit biseparable states and then illustrate our result with examples. In sec. IV, we give the concluding remark.

\section{Detection of a particular form of three-qubit biseparable state}
In this section, we will derive the coherence-based inequality which can be used to detect the given three-qubit state as biseparable states and separable states. To accomplish this task, we first find out the formula for the $l_{1}-$ norm of coherence of the tensor product of two quantum states, which are a subsystem of an n-partite system. Then, we use the derived formula to establish the required inequality for the detection of biseparable and separable states.
\subsection{Coherence of the tensor product of states}
The coherence can be measured by different measures such as distance measure, relative entropy of coherence, and $l_{p}$ norms. $l_1$ norm is a valid coherence monotone and serves as a useful measure of coherence \cite{baumgratz}. In this work, we will use the $l_1$ norm of coherence which is defined as
\begin{eqnarray}
C_{l_1}(\rho)=\sum_{i,j,i\neq j}|\rho_{ij}|
\end{eqnarray}
where $\rho_{ij}$ denotes the complex numbers corresponding to the ij-th entry of the density matrix $\rho$. $l_1$-norm of coherence of a state depends on the choice of basis in which the given state is expressed and thus from now on, we are considering only computational basis to describe the density matrix of the given state.\\
Now, we are in a position to state the result on the $l_1$ norm of coherence of the tensor product of two quantum states which are the subsystems of an n-partite system.\\\\
\textbf{Result-1:} If the density operators $\rho_{A_1,A_2,...,A_M}$ and $\rho_{A_{M+1},A_{M+2},...,A_N}$ denote the subsystem of a n-partite system, then the $l_1$ norm of coherence of the tensor product of $\rho_{A_1,A_2,...,A_M}$ and $\rho_{A_{M+1},A_{M+2},...,A_N}$ is given by
\begin{eqnarray}
&&C_{l_1}(\rho_{A_1,A_2,...,A_M}\otimes \rho_{A_{M+1},A_{M+2},...,A_N})\nonumber\\&=&C_{l_1}(\rho_{A_1,A_2,...,A_M})+C_{l_1}(\rho_{A_{M+1},A_{M+2},...,A_N})\nonumber\\&+&C_{l_1}(\rho_{A_1,A_2,...,A_M}).C_{l_1}(\rho_{A_{M+1},A_{M+2},...,A_N})
\label{result-1}
\end{eqnarray}
\textbf{Proof:}	Consider two quantum states described by the density matrices $\rho_{A_1,A_2,...,A_M}$ and $\rho_{A_{M+1},A_{M+2},...,A_N}$. The matrix representation of the density operators  $\rho_{A_1,A_2,...,A_M}$ and $\rho_{A_{M+1},A_{M+2},...A_N}$ are given by
\begin{eqnarray}
\rho_{A_1,A_2,...,A_M}=
\begin{pmatrix}
a_{1,1} & a_{1,2} & . & . & . & a_{1,m}\\
a_{2,1} & a_{2,2} & . & . & . & a_{2,m}\\
. & . & . & . & . & .\\
. & . & . & . & . & .\\
. & . & . & . & . & .\\
a_{m,1} & a_{m,2} & . & . & . & a_{m,m}
\end{pmatrix}
\end{eqnarray}
where, ~~$a_{1,1} +a_{2,2}+.+.+.+a_{m,m}=1$,
\begin{eqnarray}
\rho_{A_{M+1},A_{M+2},...A_N}=
\begin{pmatrix}
a_{m+1,m+1} & a_{m+1,m+2} & . & . & . & a_{m+1,n}\\
a_{m+2,m+1} & a_{m+2,m+2} & . & . & . & a_{m+2,n}\\
. & . & . & . & . & .\\
. & . & . & . & . & .\\
. & . & . & . & . & .\\
a_{n,m+1} & a_{n,m+2} & . & . & . & a_{n,n}
\end{pmatrix}
\end{eqnarray}
where, ~~$a_{m+1,m+1}+a_{m+2,m+2}+...+a_{n,n}=1$ and $a_{j,i}$ denotes the complex conjugate of $a_{i,j}$.\\
Then, the $l_1$ norm of coherence of the density matrices $\rho_{A_1,A_2,...,A_M}$ and $\rho_{A_{M+1},A_{M+2},...,A_N}$ are given by
\begin{eqnarray}
C_{l_1}(A_1,A_2,...,A_M)&=&\sum_{{i,j=1},{i\neq j}}^{m}{|a_{i,j}|}
\label{norm1111}
\end{eqnarray}
and
\begin{eqnarray}
C_{l_1}(\rho_{A_{M+1},A_{M+2},...,A_N})&=&\sum_{{i,j=m+1,i\neq j}}^{n}|a_{i,j}|
\label{norm1122}
\end{eqnarray}
Then,
\begin{eqnarray}
&&C_{l_1}(\rho_{A_1,A_2,...,A_M}\otimes \rho_{A_{M+1},A_{M+2},...,A_N})\nonumber\\&=&|a_{1,1}|(\sum_{{i,j=m+1},{i\neq  j}}^{n}{|a_{i,j}|})+|a_{1,2}|(\sum_{i,j=m+1}^{n}{|a_{i,j}|})\nonumber\\&+&...+|a_{1,m}|(\sum_{i,j=m+1}^{n}{|a_{i,j}|})+|a_{2,1}|(\sum_{i,j=m+1}^{n}|a_{i,j}|)\nonumber\\&+&|a_{2,2}|(\sum_{{i,j=m+1},{i\neq j}}^{n}|a_{i,j}|)+...+|a_{2,n}|(\sum_{i,j=m+1}^{n}|a_{i,j}|)\nonumber\\&+&...+|a_{m,1}|(\sum_{i,j=m+1}^{n}{|a_{i,j}|})+...\nonumber\\&+&|a_{m,m}|(\sum_{{i,j=m+1},{i\neq j}}^{n}{|a_{i,j}|})
\end{eqnarray}
Simplifying the above equation, we get
\begin{eqnarray}
&&C_{l_1}(\rho_{A_1,A_2,...,A_M}\otimes \rho_{A_{M+1},A_{M+2},...,A_N})\nonumber\\&=&\sum_{{i,j=m+1},{i\neq j}}^{n}{|a_{i,j}|}[|a_{1,1}|+|a_{2,2}|+...+|a_{m,m}|]\nonumber\\&+&\sum_{i,j=m+1}^{n}{|a_{i,j}|}[|a_{1,2}|+|a_{1,3}|+...+|a_{1,m}|+|a_{2,1}|\nonumber\\&+&|a_{2,3}|+...+|a_{2,n}|+...+|a_{m,1}|+...+|a_{m,m-1}|]
\end{eqnarray}

Using normalization condition of $\rho_{A_1,A_2,...,A_M}$ and $\rho_{A_{M+1},A_{M+2},...,A_N}$, we get
\begin{eqnarray}
&&C_{l_1}(\rho_{A_1,A_2,...,A_M}\otimes \rho_{A_{M+1},A_{M+2},...,A_N})\nonumber\\&=&\sum_{{i,j=m+1},{i\neq j}}^{n}{|a_{i,j}|}+[\sum_{i,j=m+1}^{n}{|a_{i,j}|}].[\sum_{{i,j=m+1},{i\neq j}}^{n}{|a_{i,j}|}]\nonumber\\&=&\sum_{{i,j=m+1},{i\neq j}}^{n}{|a_{i,j}|}+[1+\sum_{{i,j=m+1},{i\neq j}}^{n}{|a_{i,j}|}]\times\nonumber\\&&[\sum_{{i,j=m+1},{i\neq j}}^{n}{|a_{i,j}|}]\nonumber\\
\end{eqnarray}
From equations (\ref{norm1111}) and (\ref{norm1122}), we get,
\begin{eqnarray}
&&C_{l_1}(\rho_{A_1,A_2,...,A_M}\otimes \rho_{A_{M+1},A_{M+2},...,A_N})=C_{l_1}(\rho_{A_1,A_2,...,A_M})\nonumber\\&+&[1+C_{l_1}(\rho_{A_{M+1},A_{M+2},...,A_N})].[C_{l_1}(\rho_{A_1,A_2,...,A_M})]\nonumber
\end{eqnarray}
Thus, the $l_1$ norm of coherence of the tensor product of an m-qubit and an (n-m) qubit, which are a part of an n-partite quantum system, is given by
\begin{eqnarray}
&&C_{l_1}(\rho_{A_1,A_2,...,A_M}\otimes \rho_{A_{M+1},A_{M+2},...,A_N})\nonumber\\&=&C_{l_1}(\rho_{A_1,A_2,...,A_M})+C_{l_1}(\rho_{A_{M+1},A_{M+2},...,A_N})\nonumber\\&+&C_{l_1}(\rho_{A_1,A_2,...,A_M}).C_{l_1}(\rho_{A_{M+1},A_{M+2},...,A_N})
\end{eqnarray}
\textbf{Corollary-1:} For any two single qubit quantum states described by the density operators $\rho_1$ and $\rho_2$, the $l_1$ norm of coherence of the tensor product of $\rho_1$ and $\rho_2$ is given by
\begin{eqnarray}
C_{l_1}(\rho_1 \otimes \rho_2)=C_{l_1}(\rho_1)+C_{l_1}(\rho_2)+C_{l_1}(\rho_1).C_{l_1}(\rho_2)
\label{result-1}
\end{eqnarray}
\textbf{Proof:} Any two single-qubit quantum states described by the density matrices $\rho_1$ and $\rho_2$ is given by
\begin{eqnarray}
\rho_1=
\begin{pmatrix}
a_1 & b_1\\
b_1^{*} & d_1
\end{pmatrix}, a_1+d_1=1
\end{eqnarray}
\begin{eqnarray}
\rho_2=
\begin{pmatrix}
a_2 & b_2\\
b_2^{*} & d_2
\end{pmatrix}, a_2+d_2=1
\end{eqnarray}
The $l_1$ norm of coherence of the density matrices $\rho_1$ and $\rho_2$ are given by
\begin{eqnarray}
C_{l_1}(\rho_1)=2|b_{1}|
\label{norm1}
\end{eqnarray}
\begin{eqnarray}
C_{l_1}(\rho_2)=2|b_{2}|
\label{norm2}
\end{eqnarray}
The tensor product of $\rho_1$ and $\rho_2$ may be defined as
\begin{eqnarray}
\rho_1 \otimes \rho_2&=&
\begin{pmatrix}
a_1 & b_1\\
b_1^{*} & d_1
\end{pmatrix}
\otimes
\begin{pmatrix}
a_2 & b_2\\
b_2^{*} & d_2
\end{pmatrix}\nonumber\\&=&\begin{pmatrix}
a_{1}a_{2} & a_{1}b_{2} & b_{1}a_{2} & b_{1}b_{2}\\
a_{1}b_{2}^{*} & a_{1}d_{2} & b_{1}b_{2}^{*} & b_{1}d_{2}\\
b_{1}^{*}a_{2} & b_{1}^{*}b_{2} & d_{1}a_{2} & d_{1}b_{2}\\
b_{1}^{*}b_{2}^{*} & b_{1}^{*}d_{2} & d_{1}b_{2}^{*} & d_{1}d_{2}
\end{pmatrix}
\end{eqnarray}
The $l_1$ norm of coherence of the tensor product $\rho_1 \otimes \rho_2$ is given by
\begin{eqnarray}
C_{l_1}(\rho_1 \otimes \rho_2)&=& a_{1}|b_{2}|+|b_{1}|a_{2}+|b_{1}||b_{2}|+a_{1}|b_{2}|\nonumber\\&+&|b_{1}||b_{2}|+
|b_{1}|d_{2}+|b_{1}|a_{2}+|b_{1}||b_{2}|\nonumber\\&+&d_{1}|b_{2}|+|b_{1}||b_{2}|+|b_{1}|d_{2}+d_{1}|b_{2}|
\nonumber\\&=&2(a_{1}+d_{1})|b_2|
+2(a_{2}+d_{2})|b_1|+4|b_{1}|.|b_2|\nonumber\\&=&2|b_{1}|+2|b_{2}|+4|b_{1}|.|b_{2}|
\end{eqnarray}
From equation (\ref{norm1}) and (\ref{norm2}), we get
\begin{eqnarray}
C_{l_1}(\rho_1 \otimes \rho_2)=C_{l_1}(\rho_1)+C_{l_1}(\rho_2)+C_{l_1}(\rho_1).C_{l_1}(\rho_2)
\end{eqnarray}
Hence proved.
\\
\textbf{Corollary-2:} If the  three-qubit biseparable system described either by the density operator $\rho_{A-BC}\equiv \rho_{A}\otimes \rho_{BC}$ or $\rho_{B-AC}\equiv \rho_{B}\otimes \rho_{AC}$ or $\rho_{C-AB}\equiv \rho_{C}\otimes \rho_{AB}$, then the $l_{1}$ norm of coherence for the density operator $\rho_{A-BC}$, $\rho_{B-AC}$ and $\rho_{C-AB}$ are given by
\begin{eqnarray}
C_{l_1}(\rho_{A-BC})=C_{l_1}(\rho_A)+C_{l_1}(\rho_{BC})+C_{l_1}(\rho_A).C_{l_1}(\rho_{BC})
\label{coherence1}
\end{eqnarray}
\begin{eqnarray}
C_{l_1}(\rho_{B-AC})=C_{l_1}(\rho_{B})+C_{l_1}(\rho_{AC})+C_{l_1}(\rho_{B}).C_{l_1}(\rho_{AC})
\label{coherence2}
\end{eqnarray}
\begin{eqnarray}
C_{l_1}(\rho_{C-AB})=C_{l_1}(\rho_{C})+C_{l_1}(\rho_{AB})+C_{l_1}(\rho_{C}).C_{l_1}(\rho_{AB})
\label{coherence3}
\end{eqnarray}
\begin{enumerate}
	\item If the equality given by (\ref{coherence1}) is violated by any three-qubit state, then the state under investigation is not a biseparable state under the bipartition A-BC.
	\item If the equality given by (\ref{coherence2}) is violated by any three-qubit state, then the state under investigation is not a biseparable state of the form $\rho_{B} \otimes \rho_{AC}$.
	\item If any three-qubit state does not satisfy the equality given by (\ref{coherence3}), then the given state is not a biseparable state under the bipartition C-AB.
\end{enumerate}

\textbf{Corollary-3:} If the three-qubit system represents the separable system described by the density operator $\rho_{A-B-C}\equiv \rho_{A}\otimes \rho_{B}\otimes \rho_{C}$, then the $l_{1}$ norm of coherence for the density operator $\rho_{A-B-C}$ is given by
\begin{eqnarray}
C_{l_1}(\rho_{A-B-C})&=&C_{l_1}(\rho_A)+C_{l_1}(\rho_{B})+C_{l_1}(\rho_C)\nonumber\\&+&C_{l_1}(\rho_A).C_{l_1}(\rho_{B})+
C_{l_1}(\rho_A).C_{l_1}(\rho_{C})\nonumber\\&+&C_{l_1}(\rho_B).C_{l_1}(\rho_{C})\nonumber\\&+&C_{l_1}(\rho_A).C_{l_1}(\rho_{B}).C_{l_1}(\rho_{C})
\label{coherence4}
\end{eqnarray}
If the equality given by (\ref{coherence4}) is violated by any three-qubit state, then the state under probe is not a separable state.
\subsection{Coherence-based inequality for the detection of three-qubit biseparable states}
In this subsection, we deduce coherence-based inequality for the detection of a three-qubit biseparable state of the form $\rho^{i-jk} (i\neq j\neq k; i,j,k=A, B, C)$. To verify this inequality, we need only the information on the density matrix elements of the given three-qubit system under investigation.\\
\textbf{Result-2:} If the three-qubit state described by the density operator $\rho^{i-jk} (i\neq j\neq k; i,j,k=A, B, C)$ is biseparable such that the $l_1$-norm of coherence of at least one of the reduced system is non zero, then the $l_{1}$ norm of coherence of the biseparable state satisfies
\begin{eqnarray}
C_{l_1}(\rho^{A-BC})\leq \sum_{i}{p_i}(\frac{X_{i}^2}{4}+X_i)=U
\label{result2}
\end{eqnarray}
where 
\begin{eqnarray}
X_{i}=C_{l_1}({\rho_A^{i}})+C_{l_1}({\rho_{BC}^{i}}),~~ i=1,2,3,...
\end{eqnarray}

\textbf{Proof:} Let us consider a biseparable state for $A-BC$ partition. The biseparable state in this partition is given by
\begin{eqnarray}
\rho^{A-BC}=\sum_{i}{p_i}{\rho_A^{i}\otimes \rho_{BC}^{i}}
\end{eqnarray}
Then, $l_1$ norm of coherence of $\rho^{A-BC}$ is given by
\begin{eqnarray}
C_{l_1}({\rho^{A-BC}})&=&C_{l_1}{\sum_{i}{p_i}{\rho_A^{i}\otimes \rho_{BC}^{i}}}\nonumber\\
&\leq&\sum_{i}{p_i}(C_{l_1}({\rho_A^{i}})+C_{l_1}({\rho_{BC}^{i}})\nonumber\\&+&C_{l_1}({\rho_A^{i}}).C_{l_1}({\rho_{BC}^{i}}))
\label{eq1}
\end{eqnarray}
Now, the Arithmetic mean (AM) and Geometric mean (GM) of $C_{l_1}({\rho_A^{i}})$ and $C_{l_1}({\rho_{BC}^{i}})$ is given by
\begin{eqnarray}
\frac{C_{l_1}({\rho_A^{i}})+C_{l_1}({\rho_{BC}^{i}})}{2} \text{and} \sqrt{C_{l_1}({\rho_A^{i}}).C_{l_1}({\rho_{BC}^{i}})}
\label{am-gm}
\end{eqnarray}
respectively.\\
Using AM-GM inequality\cite{horn} on $C_{l_1}({\rho_A^{i}})$ and $C_{l_1}({\rho_{BC}^{i}})$, we get
\begin{eqnarray}
\frac{(C_{l_1}({\rho_A^{i}})+C_{l_1}({\rho_{BC}^{i}}))^2}{4}\geq C_{l_1}({\rho_A^{i}}).C_{l_1}({\rho_{BC}^{i}})
\end{eqnarray}
From (\ref{eq1}), we get
\begin{eqnarray}
C_{l_1}(\rho^{A-BC})&\leq& \sum_{i}{p_i}(C_{l_1}({\rho_A^{i}})+C_{l_1}({\rho_{BC}^{i}})\nonumber\\&+&\frac{(C_{l_1}({\rho_A^{i}})+C_{l_1}({\rho_{BC}^{i}}))^2}{4})
\label{eq2}
\end{eqnarray}
Considering $X_{i}=C_{l_1}({\rho_A^{i}})+C_{l_1}({\rho_{BC}^{i}})$ for i=1,2,3,... and using equation (\ref{eq2}), we get
\begin{eqnarray}
C_{l_1}(\rho^{A-BC})\leq \sum_{i}{p_i}(\frac{X_{i}^2}{4}+X_i)
\label{result2proof}
\end{eqnarray}
Hence proved.

\subsection{Example}
\noindent \textbf{Example-1:} Let us consider a biseparable state described by the density operator $\rho=|\psi\rangle \langle \psi|$, where $|\psi\rangle$ is given by
\begin{eqnarray}
|\psi\rangle_{ABC}=\lambda_{0}|101\rangle+\lambda_{1}|110\rangle+\lambda_{2}|111\rangle
\label{ex11}
\end{eqnarray}
where $\lambda_{0},\lambda_{1},\lambda_{2}\in R$ and $\lambda_{0}^{2}+\lambda_{1}^{2}+\lambda_{2}^{2}=1$.
Let us assume that $\lambda_{0}\geq\lambda_{1}\geq\lambda_{2}$.\\
The value of $C_{l_1}(\rho_A)$ and $C_{l_1}(\rho_{BC})$ is given by
\begin{eqnarray}
&&C_{l_1}(\rho_A)=0 \nonumber\\&&
C_{l_1}(\rho_{BC})=2(\lambda_0\lambda_1+\lambda_1\lambda_2+\lambda_0\lambda_2)
\label{coh1}
\end{eqnarray}
For the state $|\psi\rangle_{ABC}$ given in $(\ref{ex11})$, we can calculate $C_{l_1}(\rho^{A-BC})$ as
\begin{eqnarray}
C_{l_1}(\rho^{A-BC})=2(\lambda_0\lambda_1+\lambda_1\lambda_2+\lambda_0\lambda_2)
\label{exp1}
\end{eqnarray}
Using (\ref{coh1}) and (\ref{exp1}), it can be shown that the inequality (\ref{result2proof}) is satisfied.\\
\textbf{Example-2:} Consider a $|W\rangle_{ABC}$ state of the form
\begin{eqnarray}
|W\rangle_{ABC}=\frac{1}{\sqrt{3}}(|100\rangle_{ABC}+|010\rangle_{ABC}+|001\rangle_{ABC})
\label{w1}
\end{eqnarray}
The $l_{1}$ norm of coherence of the $|W\rangle_{ABC}$ state is given by $C_{l_{1}}(|W\rangle_{ABC})=2$.\\
The reduced single qubit state may be expressed as
\begin{eqnarray}
\rho^{W}_{A}=Tr_{BC}(|W\rangle_{ABC}\langle W|)=\frac{1}{3}(2|0\rangle_{A}\langle0|+|1\rangle_{A}\langle1|)
\label{w2}
\end{eqnarray}
\begin{eqnarray}
\rho^{W}_{B}=Tr_{AC}(|W\rangle_{ABC}\langle W|)=\frac{1}{3}(2|0\rangle_{B}\langle0|+|1\rangle_{B}\langle1|)
\label{w3}
\end{eqnarray}
\begin{eqnarray}
\rho^{W}_{C}=Tr_{AB}(|W\rangle_{ABC}\langle W|)=\frac{1}{3}(2|0\rangle_{C}\langle0|+|1\rangle_{C}\langle1|)
\label{w4}
\end{eqnarray}
Since the single qubit density operators $\rho^{W}_{A}$, $\rho^{W}_{B}$ and $\rho^{W}_{C}$ do not contain any off-diagonal elements so the $l_{1}$ norm of coherence for these single qubit states is given by $C_{l_{1}}(\rho^{W}_{A})=C_{l_{1}}(\rho^{W}_{B})=C_{l_{1}}(\rho^{W}_{C})=0$. Therefore, it can be easily shown that the inequality condition given in (\ref{result2}) is not maintained for the state (\ref{w1}). Further, it can be shown that the set of equality conditions given by (\ref{coherence1}), (\ref{coherence2}), and (\ref{coherence3}) are not satisfied by the state (\ref{w1}). Thus, the state (\ref{w1}) is neither a separable nor a biseparable state of the form $\rho^{W}_{A}\otimes\rho^{W}_{BC}$ or $\rho^{W}_{B}\otimes\rho^{W}_{CA}$ or $\rho^{W}_{C}\otimes\rho^{W}_{AB}$.\\
\textbf{Example 3:} Consider a $|GHZ\rangle_{ABC}$ state of the form
\begin{eqnarray}
|GHZ\rangle_{ABC}=Cos\theta|000\rangle+e^{i\delta}Sin\theta|111\rangle
\label{ghz1}
\end{eqnarray}
The $l_{1}$ norm of coherence of the $|GHZ\rangle_{ABC}$ state is given by $C_{l_{1}}(|GHZ\rangle_{ABC})=2e^{i\delta}Sin\theta Cos\theta$.\\
The reduced single qubit state may be expressed as
\begin{eqnarray}
\rho^{GHZ}_{A}&=&Tr_{BC}(|GHZ\rangle_{ABC}\langle GHZ|)\nonumber\\&=&Cos^{2}\theta|0\rangle_{A}\langle0|+e^{2i\delta}Sin^{2}\theta|1\rangle_{A}\langle1|)
\label{ghz2}
\end{eqnarray}
\begin{eqnarray}
\rho^{GHZ}_{B}&=&Tr_{AC}(|GHZ\rangle_{ABC}\langle GHZ|)\nonumber\\&=&Cos^{2}\theta|0\rangle_{B}\langle0|+e^{2i\delta}Sin^{2}\theta|1\rangle_{B}\langle1|)
\label{ghz3}
\end{eqnarray}
\begin{eqnarray}
\rho^{GHZ}_{C}&=&Tr_{AB}(|GHZ\rangle_{ABC}\langle GHZ|)\nonumber\\&=&Cos^{2}\theta|0\rangle_{C}\langle0|+e^{2i\delta}Sin^{2}\theta|1\rangle_{C}\langle1|)
\label{ghz4}
\end{eqnarray}
Since the single qubit density operators $\rho^{GHZ}_{A}$, $\rho^{GHZ}_{B}$ and $\rho^{GHZ}_{C}$ do not contain any off-diagonal elements so the $l_{1}$ norm of coherence for these single qubit states is given by $C_{l_{1}}(\rho^{GHZ}_{A})=C_{l_{1}}(\rho^{GHZ}_{B})=C_{l_{1}}(\rho^{GHZ}_{C})=0$. 
Also, the reduced two-qubit system may be expressed as
\begin{eqnarray}
\rho^{GHZ}_{AB}&=&Tr_{C}(|GHZ\rangle_{ABC}\langle GHZ|)\nonumber\\&=&Cos^{2}\theta|00\rangle_{A}\langle00|+e^{2i\delta}Sin^{2}\theta|11\rangle_{A}\langle11|
\label{ghz2}
\end{eqnarray}
\begin{eqnarray}
\rho^{GHZ}_{AC}&=&Tr_{B}(|GHZ\rangle_{ABC}\langle GHZ|)\nonumber\\&=&Cos^{2}\theta|00\rangle_{B}\langle00|+e^{2i\delta}Sin^{2}\theta|11\rangle_{B}\langle11|
\label{ghz3}
\end{eqnarray}
\begin{eqnarray}
\rho^{GHZ}_{BC}&=&Tr_{A}(|GHZ\rangle_{ABC}\langle GHZ|)\nonumber\\&=&Cos^{2}\theta|00\rangle_{C}\langle00|+e^{2i\delta}Sin^{2}\theta|11\rangle_{C}\langle11|
\label{ghz4}
\end{eqnarray}
Since the reduced two-qubit density operators $\rho^{GHZ}_{AB}$, $\rho^{GHZ}_{BC}$ and $\rho^{GHZ}_{AC}$ does not contain any off-diagonal elements so the $l_{1}$ norm of coherence for these two-qubit states is given by $C_{l_{1}}(\rho^{GHZ}_{AB})=C_{l_{1}}(\rho^{GHZ}_{BC})=C_{l_{1}}(\rho^{GHZ}_{AC})=0$.\\
Therefore, it can be easily seen that the inequality is given in (\ref{result2}) is not maintained for the state (\ref{ghz1}). Further, it can be shown that the set of equality conditions given by (\ref{coherence1}), (\ref{coherence2}), and (\ref{coherence3}) are not satisfied by the state (\ref{w1}). Thus, the state (\ref{ghz1}) is neither a separable nor a biseparable state of the form $\rho^{GHZ}_{A}\otimes\rho^{GHZ}_{BC}$ or $\rho^{GHZ}_{B}\otimes\rho^{GHZ}_{CA}$ or $\rho^{GHZ}_{C}\otimes\rho^{GHZ}_{AB}$.\\

\textbf{Example 4:} Consider a state described by the density operator $\rho^{(1)}_{ABC}=|\psi\rangle_{ABC}\langle \psi|$, where $|\psi\rangle_{ABC}$ is given by
\begin{eqnarray}
|\psi\rangle_{ABC}=a_0|000\rangle+a_1|100\rangle+\frac{1}{\sqrt{2}}|111\rangle,~~a_0,a_1\geq 0
\label{ghz5}
\end{eqnarray}
The normalization condition of the state $|\psi\rangle_{ABC}$ gives
\begin{eqnarray}
a_{0}^{2}+a_{1}^{2}=\frac{1}{2}
	\label{nor5}
\end{eqnarray}
The $l_{1}$-norm of coherence of the $\rho^{(1)}_{ABC}$ state is given by $C_{l_{1}}(\rho^{(1)}_{ABC})=2a_0a_1+\sqrt{2}a_0+\sqrt{2}a_1$.\\
The reduced single qubit state may be expressed as
\begin{eqnarray}
\rho^{(1)}_{A}&=&Tr_{BC}(|\psi \rangle_{ABC}\langle \psi|)\nonumber\\&=&a_0^2|0\rangle \langle 0|+2a_0a_1|1\rangle \langle 0|+({a_1}^{2}+\frac{1}{2})|1\rangle \langle 1|\\
\rho^{(1)}_{B}&=&Tr_{AC}(|\psi \rangle_{ABC}\langle \psi|)\nonumber\\&=&(a_0^2+{a_1}^{2})|0\rangle \langle 0|+\frac{1}{2}|1\rangle \langle 1|\\
\rho^{(1)}_{C}&=&Tr_{AB}(|\psi \rangle_{ABC}\langle \psi|)\nonumber\\&=&(a_0^2+{a_1}^{2})|0\rangle \langle 0|+\frac{1}{2}|1\rangle \langle 1|
\end{eqnarray}
The $l_1$-norm of coherence of the single qubit density operators $\rho^{(1)}_A$, $\rho^{(1)}_B$ and $\rho^{(1)}_C$ may be expressed as
\begin{eqnarray}
C_{l_1}(\rho^{(1)}_A)=2a_0a_1,~
C_{l_1}(\rho^{(1)}_B)=C_{l_1}(\rho^{(1)}_C)=0
\end{eqnarray}
and the $l_1$-norm of coherence of the reduced two-qubit system may be expressed as
\begin{eqnarray}
C_{l_1}(\rho^{(1)}_{AB})=C_{l_1}(\rho^{(1)}_{AC})=2a_0a_1,C_{l_1}(\rho^{(1)}_{BC})=\sqrt{2}{a_1}
\end{eqnarray}  
\begin{figure}[h]
	\centering
	\includegraphics[scale=.6]{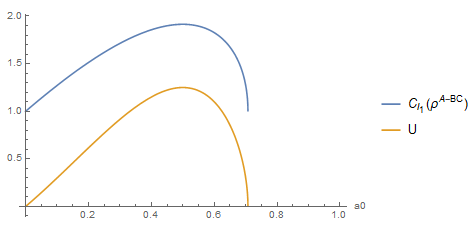}
	\caption{Comparision between the $l_1$-norm of coherence of the state $\rho^{(1)}$ and the upper bound "U" obtained in Result-2.}
\end{figure} 
From fig-1, it can be seen that the inequality given in Result-2 is violated for the state $\rho^{(1)}$ for the bipartition cut $A-BC$. Similarly, it can be seen that Result-2 is violated for the state $\rho^{(1)}$ in the bipartition cuts $B-AC$ and $C-AB$. Thus the state described by the density operator $\rho^{(1)}$ is not a separable state. Further, it can be shown that the set of equality conditions given by (\ref{coherence1}), (\ref{coherence2}), and (\ref{coherence3}) are not satisfied by the state (\ref{ghz5}). Thus, the state (\ref{ghz5}) is neither a separable nor a biseparable state of the form $\rho^{(1)}_{A}\otimes\rho^{(1)}_{BC}$ or $\rho^{(1)}_{B}\otimes\rho^{(1)}_{CA}$ or $\rho^{(1)}_{C}\otimes\rho^{(1)}_{AB}$.

\section{Detection of general three-qubit biseparable states and separable states}
A mixed state is said to be fully separable if it can be written as the convex combination of fully separable pure states.
A mixed state is said to be biseparable if it is not fully separable and it can be written as a convex combination of biseparable pure states.
Let us recall the three-qubit mixed biseparable state given in (\ref{generalbiseparable}) and re-write it as
\begin{eqnarray}
\sigma_{bisep} = p_1 \sigma^{A-BC}_{bisep} + p_2 \sigma^{B-AC}_{bisep} + p_3 \sigma^{C-AB}_{bisep},\sum_{i}p_{i}=1
\label{bisep11}
\end{eqnarray}
where, $0\leq p_i\leq 1$.
\subsection{Coherence-based inequality for the detection of general three-qubit biseparable states}
Detection of three-qubit mixed states has been studied by constructing the witness operator \cite{acin}. In this section, we will study the detection of three-qubit mixed biseparable states using coherence-based inequality.\\
\textbf{Result-3:} If the three-qubit mixed state described by the density operator in (\ref{bisep11}) is biseparable then it satisfies the inequality
\begin{eqnarray}
1+C_{l_1}(p_1\sigma_1^{A-BC}+p_2\sigma_2^{B-CA}+p_3\sigma_3^{C-AB})\nonumber\\ \leq \frac{1}{4}\sum_{i=1}^{3}p_i(X_i+2)^2
\label{result3}
\end{eqnarray}
where,
\begin{eqnarray}
X_1=C_{l_1}(\sigma_1^{A})+C_{l_1}(\sigma_1^{BC})\nonumber\\
X_2=C_{l_1}(\sigma_2^{B})+C_{l_1}(\sigma_2^{CA})\nonumber\\
X_3=C_{l_1}(\sigma_3^{C})+C_{l_1}(\sigma_3^{AB})
\label{x1x2x3}
\end{eqnarray}

\textbf{Proof:} Let us consider a mixed three-qubit biseparable state whose density matrix is given by $p_1\sigma_1^{A-BC}+p_2\sigma_2^{B-CA}+p_3\sigma_3^{C-AB}$. Using $l_1$ norm of coherence of $p_1\sigma_1^{A-BC}+p_2\sigma_2^{B-CA}+p_3\sigma_3^{C-AB}$ and the convexity property of $l_1$ norm of coherence, we have,
\begin{eqnarray}
C_{l_1}(p_1\sigma_1^{A-BC}+p_2\sigma_2^{B-CA}+p_3\sigma_3^{C-AB})
\leq \nonumber\\ p_1C_{l_1}(\sigma_1^{A-BC})+p_2C_{l_1}(\sigma_2^{B-CA})+p_3C_{l_1}(\sigma_3^{C-AB})
\end{eqnarray}
Using the relation (\ref{coherence1}) or other related coherence relation like (\ref{coherence2}) or (\ref{coherence3}), we get
\begin{eqnarray}
&&C_{l_1}(p_1\sigma_1^{A-BC}+p_2\sigma_2^{B-CA}+p_3\sigma_3^{C-AB}) \nonumber\\&\leq& p_1[C_{l_1}(\sigma_1^{A}+C_{l_1}(\sigma_1^{BC})+C_{l_1}(\sigma_1^{A}).C_{l_1}(\sigma_1^{BC})]\nonumber\\&+&p_2[C_{l_1}(\sigma_2^{B}+C_{l_1}(\sigma_2^{CA})+C_{l_1}(\sigma_2^{B}).C_{l_1}(\sigma_2^{CA})]\nonumber\\&+&p_3[C_{l_1}(\sigma_3^{C}+C_{l_1}(\sigma_3^{AB})+C_{l_1}(\sigma_3^{C}).C_{l_1}(\sigma_3^{AB})]
\label{ineq43}
\end{eqnarray}
Recalling expressions of AM and GM of $C_{l_1}(\sigma_{i}^{A})$ and $C_{l_1}(\sigma_{i}^{BC})$ from (\ref{am-gm}) and applying $AM\geq GM$ on $C_{l_1}(\sigma_{i}^{A})$ and $C_{l_1}(\sigma_{i}^{BC})$, inequality (\ref{ineq43}) reduces to,
\begin{eqnarray}
&&C_{l_1}(p_1\sigma_1^{A-BC}+p_2\sigma_2^{B-CA}+p_3\sigma_3^{C-AB}) \nonumber\\&\leq& p_1[C_{l_1}(\sigma_1^{A})+C_{l_1}(\sigma_1^{BC})+\frac{1}{4}(C_{l_1}(\sigma_1^{A})+C_{l_1}(\sigma_1^{BC}))^2]\nonumber\\&+&p_2[C_{l_1}(\sigma_2^{B})+C_{l_1}(\sigma_2^{CA})+\frac{1}{4}(C_{l_1}(\sigma_2^{B})+C_{l_1}(\sigma_2^{CA}))^2]\nonumber\\&+&p_3[C_{l_1}(\sigma_3^{C})+C_{l_1}(\sigma_3^{AB})+\frac{1}{4}(C_{l_1}(\sigma_3^{C})+C_{l_1}(\sigma_3^{AB}))^2]
\end{eqnarray}
Thus, we have
\begin{eqnarray}
&&C_{l_1}(p_1\sigma_1^{A-BC}+p_2\sigma_2^{B-CA}+p_3\sigma_3^{C-AB}) \nonumber\\&\leq&p_1(X_1+\frac{X_1^{2}}{4})+p_2(X_2+\frac{X_2^{2}}{4})+p_3(X_3+\frac{X_3^{2}}{4})
\end{eqnarray}

Therefore,
\begin{eqnarray}
&&C_{l_1}(p_1\sigma_1^{A-BC}+p_2\sigma_2^{B-CA}+p_3\sigma_3^{C-AB}) \nonumber\\&\leq&p_1(X_1+\frac{X_1^{2}}{4})+p_2(X_2+\frac{X_2^{2}}{4})+p_3(X_3+\frac{X_3^{2}}{4})\nonumber\\&=&\frac{1}{4}[p_1(X_1+2)^2+p_2(X_2+2)^2+p_3(X_3+2)^2]-1\nonumber\\
\label{ineq8}
\end{eqnarray}
where $X_{1}, X_2$ and $X_3$ are given by (\ref{x1x2x3}).\\
Simplifying (\ref{ineq8}), we get
\begin{eqnarray}
1+C_{l_1}(p_1\sigma_1^{A-BC}+p_2\sigma_2^{B-CA}+p_3\sigma_3^{C-AB})\nonumber\\ \leq \frac{1}{4}\sum_{i=1}^{3}p_i(X_i+2)^2
\end{eqnarray}
which is the required result. Hence proved.\\
\\
\textbf{Corollary-4:} If any three-qubit mixed state violates the inequality (\ref{result3}), then the given state is not a biseparable state.

\subsection{Coherence-based inequality for the detection of three-qubit mixed separable states}
In this subsection, we will study the detection of three-qubit mixed separable states using the inequality based on the coherence of a single qubit. Let us consider a mixed separable state, which can be expressed as
\begin{eqnarray}
\sigma^{A-B-C}=\sum_{i}p_{i}\sigma_{i}^{A}\otimes\sigma_{i}^{B}\otimes\sigma_{i}^{C}
\label{sepmix100}
\end{eqnarray}
The $l_{1}$ norm of the coherence of the state (\ref{sepmix100}) is given by
\begin{eqnarray}
C_{l_1}(\sigma^{A-B-C})&=& C_{l_{1}}(\sum_{i}p_{i}\sigma_{i}^{A}\otimes\sigma_{i}^{B}\otimes\sigma_{i}^{C})\nonumber\\
&\leq& \sum_{i}p_{i}C_{l_{1}}(\sigma_{i}^{A}\otimes\sigma_{i}^{B}\otimes\sigma_{i}^{C})\nonumber\\&=& \sum_{i}p_{i}[\sum_{x=A,B,C}C_{l_{1}}(\sigma_{i}^{x})\nonumber\\&+&\sum_{x\neq y,x,y=A,B,C}C_{l_{1}}(\sigma_{i}^{x})C_{l_{1}}(\sigma_{i}^{y})\nonumber\\&+&C_{l_{1}}(\sigma_{i}^{A})C_{l_{1}}(\sigma_{i}^{B})C_{l_{1}}(\sigma_{i}^{C})]
\end{eqnarray}
The inequality in the second step follows from the convexity property of $l_{1}$ norm of coherence. Therefore, we are now in a position to state the
result of the derived inequality for the mixed three-qubit separable state.\\
\textbf{Result-4:} If the three-qubit mixed state described by the density operator $\sigma^{A-B-C}=\sum_{i}p_{i}\sigma_{i}^{A}\otimes\sigma_{i}^{B}\otimes\sigma_{i}^{C}$ is separable then it satisfies the inequality
\begin{eqnarray}
C_{l_1}(\sigma^{A-B-C})&\leq& \sum_{i}p_{i}[\sum_{x=A,B,C}C_{l_{1}}(\sigma_{i}^{x})\nonumber\\&+&\sum_{x\neq y,x,y=A,B,C}C_{l_{1}}(\sigma_{i}^{x})C_{l_{1}}(\sigma_{i}^{y})\nonumber\\&+&C_{l_{1}}(\sigma_{i}^{A})C_{l_{1}}(\sigma_{i}^{B})C_{l_{1}}(\sigma_{i}^{C})]
\label{sepmix2000}
\end{eqnarray}
\textbf{Corollary-5:} If any three-qubit mixed state violates the inequality (\ref{sepmix2000}) then the given state is not a separable state.
\subsection{Illustrations}
\textbf{Example-1:} Consider a mixed three-qubit biseparable state described by the density matrix $\rho_1$, which is given by
\begin{eqnarray}
\rho_1&=&q|0\rangle^{A} \langle 0|) \otimes |\phi^{+}\rangle^{BC} \langle \phi^{+}|\nonumber\\&+&(1-q)|1\rangle^{B} \langle 1|\otimes |\phi^{-}\rangle^{AC} \langle \phi^{-}|
\label{ex3}
\end{eqnarray}
where $0\leq q\leq 1$ and the Bell states $|\phi^{+}\rangle^{BC}$ and $|\phi^{-}\rangle^{AC}$ are given by
\begin{eqnarray}
|\phi^{+}\rangle^{BC} &=&\frac{1}{\sqrt{2}}(|00\rangle^{BC}+|11\rangle^{BC})\nonumber\\
|\phi^{-}\rangle^{AC}&=&\frac{1}{\sqrt{2}}(|00\rangle^{AC}-|11\rangle^{AC})
\end{eqnarray}
Comparing equation (\ref{ex3}) with general mixed three-qubit biseparable state, we have $p_1=q$, $p_2=1-q$, and $p_3=0$. $l_1$ norm of coherence for the given state, defined in (\ref{ex3}), is given by
\begin{eqnarray}
C_{l_1}(\rho_{1})=1
\end{eqnarray}
For the given state $\rho_1$, $C_{l_1}(\rho_1^{A})=C_{l_1}(\rho_1^{B})=C_{l_1}(\rho_1^{C})=0$, $C_{l_1}(\rho_1^{BC})=1$, $C_{l_1}(\rho_1^{AC})=1$ and $C_{l_1}(\rho_1^{AB})=0$.
The quantity $X_{i}'s$, i=1,2,3 can be calculated as
\begin{eqnarray}
X_{1}&=&C_{l_1}(\rho_{1}^{A})+C_{l_1}(\rho_{1}^{BC})=1\nonumber\\
X_{2}&=&C_{l_1}(\rho_{1}^{B})+C_{l_1}(\rho_{1}^{AC})=1\nonumber\\
X_{3}&=&C_{l_1}(\rho_{1}^{C})+C_{l_1}(\rho_{1}^{AB})=0
\label{X1ex1}
\end{eqnarray}
Substituting values of $p_i's$, $X_i's$ and $l_1$ norm of coherence of $\rho_1$, we can see that equation (\ref{result3})
is satisfied for the state (\ref{ex3}).\\

\textbf{Example-2:} Let us consider the mixed three-qubit state described by the density operator $\varrho^{ABC}$ as
\begin{eqnarray}
\varrho^{ABC}&=&q|GHZ\rangle \langle GHZ|)+(1-q)|W\rangle \langle W|
\label{mixedex2}
\end{eqnarray}
where $0\leq q\leq 1$ and the three-qubit states $|GHZ\rangle$ and $|W\rangle$ are given by
\begin{eqnarray}
|GHZ\rangle =\frac{1}{\sqrt{2}}(|000\rangle+|111\rangle)\\
|W\rangle =\frac{1}{\sqrt{3}}(|001\rangle+|010\rangle+|100\rangle
\end{eqnarray}
Comparing equation (\ref{mixedex2}) with general three-qubit mixed biseparable state, we get $p_1=q$, $p_2=1-q$ and $p_3=0$.\\
The $l_1$ norm of coherence of the state (\ref{mixedex2}) is given by
\begin{eqnarray}
C_{l_1}(\varrho^{ABC})=3
\label{Mex2}
\end{eqnarray}
For the given state $\varrho^{ABC}$, $C_{l_1}(\varrho^{A})= C_{l_1}(\varrho^{B})=C_{l_1}(\varrho^{C})=0$ and $ C_{l_1}(\varrho^{BC})=C_{l_1}(\varrho^{AC})=C_{l_1}(\varrho^{AB})=\frac{2}{3}$. Therefore, the quantities $X_i's$, i=1,2,3 can be written as,
\begin{eqnarray}
X_{1}&=&C_{l_1}(\rho_{1}^{A})+C_{l_1}(\rho_{1}^{BC})=\frac{2}{3}\nonumber\\
X_{2}&=&C_{l_1}(\rho_{1}^{B})+C_{l_1}(\rho_{1}^{AC})=\frac{2}{3}\nonumber\\
X_{3}&=&C_{l_1}(\rho_{1}^{C})+C_{l_1}(\rho_{1}^{AB})=\frac{2}{3}
\end{eqnarray}
Substituting values of $p_i's$, $X_i's$ and $C_{l_1}(\varrho^{ABC})$ in equation (\ref{result3}), we can observe that the inequality given in result-3 is voilated for any q. Thus, the given state $\varrho^{ABC}$ is not a biseparable state. Moreover, a simple calculation also shows that the inequality (\ref{sepmix200}) is violated for any $q$. So we can infer that the given state (\ref{mixedex2}) is not a separable state. Thus, we find that the given state $\varrho^{ABC}$ is neither a biseparable state nor a separable state. Hence by using our criterion, we detect that the given state is a genuine mixed three-qubit entangled state.\\

\section{Conclusion}
To summarize, we have formulated the $l_{1}$ norm of coherence for the tensor product $m-qubit \otimes (n-m)-qubit$. 
Using the derived formula for the tensor product of two quantum states that represents two subsystems of an n-partite system, we provide the necessary conditions in terms of coherence-based inequalities for the detection of three-qubit biseparable states of a particular form. Thus, if any three-qubit state violates the corresponding coherence-based inequality then the given three-qubit state is definitely not a biseparable state of the form $\rho_{A}\otimes \rho_{BC}$ or $\rho_{B}\otimes \rho_{CA}$ or $\rho_{C}\otimes \rho_{AB}$. We have also derived the necessary condition for the detection of the most general biseparable mixed state. We further apply our criterion to some specific examples of the three-qubit system to detect the nature of bipartite entanglement in a three-qubit system. For pure states, $l_1$ norm of coherence is equal to the robustness of coherenence$(C_R)$.  Robustness of coherence quantifies how much noise must be added in order to make the state separable\cite{toth2}. It can be evaluated experimentally with linear optics using (a)an interference-fringe method and (b) the witness approach\cite{wang}. Thus, the robustness of coherence can be calculated experimentally, and hence $l_1$ norm of coherence for pure states can be calculated experimentally. The results we have discussed in this work can be easily generalized to a multipartite (more than three parties) as well as a higher dimensional quantum system. It is evident from Appendix-1 and Appendix-2.

\section{Acknowledgement}
A.K. would like to acknowledge the financial support from CSIR. This work is supported by CSIR File No. 08/133(0027)/2018-EMR-1.\\

\section{Data Availability Statement}
Data sharing not applicable to this article as no datasets were generated or analysed during the current study.

\section{Appendix-1}
\noindent The results we have obtained in our work can be generalized to a multipartite system also. For instance, if we consider the four-qubit system, Result-3 for mixed biseparable states may be re-stated as:\\
If the four-qubit mixed state described by the density operator
\begin{eqnarray}
\rho=p_1\sigma_1^{A-BCD}+p_2\sigma_2^{B-CAD}+p_3\sigma_3^{C-ABC}+p_4\sigma_4^{D-ABC}
\label{bisep4}
\end{eqnarray} is biseparable then it satisfies the inequality
\begin{eqnarray}
1+C_{l_1}(p_1\sigma_1^{A-BCD}+p_2\sigma_2^{B-CAD}+p_3\sigma_3^{C-ABD}+p_4\sigma_4^{D-ABC})\nonumber\\ \leq \frac{1}{4}\sum_{i=1}^{4}p_i(X_i+2)^2
\label{result31}
\end{eqnarray}
where,
\begin{eqnarray}
X_1=C_{l_1}(\sigma_1^{A})+C_{l_1}(\sigma_1^{BCD})\nonumber\\
X_2=C_{l_1}(\sigma_2^{B})+C_{l_1}(\sigma_2^{CAD})\nonumber\\
X_3=C_{l_1}(\sigma_3^{C})+C_{l_1}(\sigma_3^{ABD})\nonumber\\
X_4=C_{l_1}(\sigma_4^{D})+C_{l_1}(\sigma_4^{ABC})
\label{x1x2x3}
\end{eqnarray}
To verify this result, let us consider a mixed biseparable state in a four-qubit system described by the density matrix $\rho_1$, which is given by
\begin{eqnarray}
\rho_1&=&\frac{1}{2}[|0\rangle^{A} \langle 0|) \otimes |\phi^{+}\rangle^{BCD} \langle \phi^{+}|]\nonumber\\&+&\frac{1}{2}[|1\rangle^{B} \langle 1|\otimes |\phi^{-}\rangle^{ACD} \langle \phi^{-}|]
\label{ex}
\end{eqnarray}
where the states $|\phi^{+}\rangle^{BCD}$ and $|\phi^{-}\rangle^{ACD}$ are given by
\begin{eqnarray}
|\phi^{+}\rangle^{BCD} &=&\frac{1}{\sqrt{2}}(|100\rangle^{BCD}+|010\rangle^{BCD})\nonumber\\
|\phi^{-}\rangle^{ACD}&=&\frac{1}{\sqrt{2}}(|100\rangle^{ACD}-|010\rangle^{ACD})
\end{eqnarray}
Comparing equation (\ref{ex}) with general mixed four qubit biseparable state (\ref{bisep4}), we have $p_1=\frac{1}{2}$, $p_2=\frac{1}{2}$, $p_3=0$ and $p_4=0$. $l_1$-norm of coherence for the given state, defined in (\ref{ex}), is given by
\begin{eqnarray}
C_{l_1}(\rho_{1})=1
\end{eqnarray}
For the given state $\rho_1$, $C_{l_1}(\rho_1^{A})=C_{l_1}(\rho_1^{B})=C_{l_1}(\rho_1^{C})=C_{l_1}(\rho_1^{D})=0$, $C_{l_1}(\rho_1^{BCD})=1$, $C_{l_1}(\rho_1^{ACD})=1$, $C_{l_1}(\rho_1^{ABD})=1$ and $C_{l_1}(\rho_1^{ABC})=0$.
The quantity $X_{i}'s$, i=1,2,3,4 can be calculated as
\begin{eqnarray}
X_{1}&=&C_{l_1}(\rho_{1}^{A})+C_{l_1}(\rho_{1}^{BCD})=1\nonumber\\
X_{2}&=&C_{l_1}(\rho_{1}^{B})+C_{l_1}(\rho_{1}^{ACD})=1\nonumber\\
X_{3}&=&C_{l_1}(\rho_{1}^{C})+C_{l_1}(\rho_{1}^{ABD})=0\nonumber\\
X_{4}&=&C_{l_1}(\rho_{1}^{D})+C_{l_1}(\rho_{1}^{ABC})=0
\label{X1ex}
\end{eqnarray}
Substituting the values of $p_i's$, $X_i's$ and $l_1$-norm of coherence of $\rho_1$ in (\ref{result31}), it can be easily seen that (\ref{result31})
is satisfied for the state defined in (\ref{ex}).\\
Also, for four-qubit mixed separable states, Result-4 may be re-stated as:\\
If the four-qubit mixed state described by the density operator $\sigma^{A-B-C}=\sum_{i}p_{i}\sigma_{i}^{A}\otimes\sigma_{i}^{B}\otimes\sigma_{i}^{C}\otimes\sigma_{i}^{D}$ is separable then it satisfies the inequality
\begin{eqnarray}
C_{l_1}(\sigma^{A-B-C-D})&\leq& \sum_{i}p_{i}[\sum_{x=A,B,C,D}C_{l_{1}}(\sigma_{i}^{x})\nonumber\\&+&\sum_{x\neq y,x,y=A,B,C,D}C_{l_{1}}(\sigma_{i}^{x})C_{l_{1}}(\sigma_{i}^{y})\nonumber\\&+&\sum_{x\neq y\neq z,x,y,z=A,B,C,D}C_{l_{1}}(\sigma_{i}^{x})C_{l_{1}}(\sigma_{i}^{y})C_{l_{1}}(\sigma_{i}^{z})\nonumber\\&+&C_{l_{1}}(\sigma_{i}^{A})C_{l_{1}}(\sigma_{i}^{B})C_{l_{1}}(\sigma_{i}^{C})]
\label{sepmix200}
\end{eqnarray}
To verify the above result for a four-qubit mixed separable state, let us consider a separable state in a four-qubit system, described by the density operator $\rho_2$,
\begin{eqnarray}
\rho_2&=&\frac{1}{4}|0000\rangle^{ABCD}\langle 0000|+\frac{1}{4}|0011\rangle^{ABCD} \langle 0011|\nonumber\\&+&\frac{1}{4}|1000\rangle^{ABCD}\langle 1000|+\frac{1}{4}|1111\rangle^{ABCD} \langle 1111|
\label{ex2}
\end{eqnarray}
Then, for the state defined in (\ref{ex2}), $C_{l_1}(\rho_2)=0$ and $C_{l_1}(\rho_2^{A})=C_{l_1}(\rho_2^{B})=C_{l_3}(\rho_2^{C})=C_{l_1}(\rho_2^{D})=0$, thus we can say that the equation (\ref{sepmix200}) is verified for the state $\rho_2$.\\

\section{Appendix-2}
\noindent The results we have obtained in our work can be generalized to a higher dimensional quantum system. For example if we consider a state $\rho=|\psi\rangle \langle \psi|$ in a three-qutrit system, where $|\psi\rangle$ is given by,
\begin{eqnarray}
|\psi\rangle=|0\rangle \otimes \frac{1}{\sqrt{3}}(|12\rangle +|01\rangle+|20\rangle)
\label{ex3}
\end{eqnarray}
Comparing equation (\ref{ex3}) with the general mixed three-qubit biseparable state of our manuscript, we have $p_1=1$, $p_2=0$, and $p_3=0$. $l_1$ norm of coherence for the given state, defined in (\ref{ex3}), is given by
\begin{eqnarray}
C_{l_1}(\rho)=2
\end{eqnarray}
For the given state $\rho$, $C_{l_1}(\rho^{A})=C_{l_1}(\rho^{B})=C_{l_1}(\rho^{C})=0$, $C_{l_1}(\rho^{BC})=2$, $C_{l_1}(\rho^{AC})=0$ and $C_{l_1}(\rho^{AB})=0$.
The quantity $X_{i}'s$, i=1,2,3 can be calculated as
\begin{eqnarray}
X_{1}&=&C_{l_1}(\rho^{A})+C_{l_1}(\rho^{BC})=2\nonumber\\
X_{2}&=&C_{l_1}(\rho^{B})+C_{l_1}(\rho^{AC})=0\nonumber\\
X_{3}&=&C_{l_1}(\rho^{C})+C_{l_1}(\rho^{AB})=0
\label{X1ex1}
\end{eqnarray}
Substituting values of $p_i's$, $X_i's$ and $l_1$ norm of coherence of $\rho$, we can see that equation (\ref{result3}) is satisfied for the state (\ref{ex3}).\\
Thus, the given state is a biseparable state. Hence, our result also holds for higher dimensional system.


\end{document}